%% file: ssp20.tex
\documentclass[conference]{IEEEtran}

\usepackage{amsmath,dsfont,bbm,epsfig,amssymb,amsfonts,amstext,verbatim,amsopn,cite}
\usepackage{subfigure,multirow,multicol,lipsum,xfrac}
\usepackage{amsthm,ulem}
\usepackage{mathtools,amsthm}
\usepackage{perpage}
\usepackage{balance}
\usepackage{url}
\usepackage{amsfonts}
\usepackage{epsfig}
\usepackage[font={small}]{caption}
\usepackage{etoolbox}
\usepackage{algorithmicx}
\usepackage[Algorithm,ruled]{algorithm}
\usepackage{algpseudocode}
\usepackage{pifont}
\usepackage[utf8]{inputenc}
\usepackage[T1]{fontenc}  
\usepackage[nolist]{acronym}
\MakePerPage{footnote}
\usepackage{textcomp}
\usepackage{paralist}
\usepackage{enumitem}
\usepackage{bbm}
\usepackage[process=auto]{pstool}
\usepackage{tikz,pgfplots}
\usetikzlibrary{shapes,arrows}

\include{commands}

\renewcommand{\emph}[1]{\textit{#1}}

\begin{document}
\title{Efficient Matrix Multiplication: The Sparse Power-of-2 Factorization}
\author{
\IEEEauthorblockN{
Ralf R. M\"uller\IEEEauthorrefmark{1},
Bernhard G\"ade\IEEEauthorrefmark{1},
Ali Bereyhi\IEEEauthorrefmark{1}
}
\IEEEauthorblockA{
\IEEEauthorrefmark{1}Institute for Digital Communications, Friedrich-Alexander Universit\"at Erlangen-N\"urnberg, Germany\\
ralf.r.mueller@fau.de, bernhard.gaede@fau.de, ali.bereyhi@fau.de
}
}
%
%
\IEEEoverridecommandlockouts
\maketitle

\begin{acronym}
\acro{oas}[OAS]{oversampled adaptive sensing}
\acro{csi}[CSI]{channel state information}
\acro{awgn}[AWGN]{additive white Gaussian noise}
\acro{iid}[i.i.d.]{independent and identically distributed}
\acro{rhs}[r.h.s.]{right hand side}
\acro{lhs}[l.h.s.]{left hand side}
\acro{wrt}[w.r.t.]{with respect to}
\acro{rs}[RS]{replica symmetry}
\acro{rsb}[RSB]{replica symmetry breaking}
\acro{mse}[MSE]{mean squared error}
\acro{mmse}[MMSE]{minimum MSE}
\acro{sinr}[SINR]{signal to interference and noise ratio}
\acro{mf}[MF]{matched filtering}
\end{acronym}
\begin{abstract}
We present an algorithm to reduce the computational effort for the multiplication of a given matrix with an unknown column vector.
The algorithm decomposes the given matrix into a product of matrices whose entries are either zero or integer powers of two utilizing the principles of sparse recovery. While classical low resolution quantization achieves an accuracy of 6 dB per bit, our method can achieve many times more than that for large matrices. Numerical and analytical evidence suggests that the improvement actually grows unboundedly with matrix size. Due to sparsity, the algorithm even allows for quantization levels below 1 bit per matrix entry while achieving highly accurate approximations for large matrices.
Applications include, but are not limited to, neural networks, as well as fully digital beam-forming for massive MIMO and millimeter wave applications.
\end{abstract}

\IEEEpeerreviewmaketitle

\section{Introduction}
\label{sec:intro}
In various applications, multiplications of a data vector with a given large matrix have to be performed at very high rates; see
\cite{Fatema2018Massive,yan2019performance,orponen1994computational,takeda1986neural,
judd1990neural,bianchini2014complexity} and the references there in for some examples. In order to reduce the power consumption of the circuitry, the matrices are often quantized to low levels of bit resolution resulting in severe distortion by quantization noise.
It is well-known from standard textbooks that any additional bit increases the signal-to-noise\footnote{Whenever we refer to noise, we always mean quantization noise.} ratio (SNR) by a factor of 4, i.e.\ 6 dB; see for example \cite{zimmermann1999computer,patterson,oppenheim}.

In Section~\ref{alg}, we present an algorithm that performs much better than 6 dB/bit for large matrices with ``large'' starting at matrices with the greater of the two dimension being at least 32. In Section~\ref{num}, we give numerical results on its performance.
In Section~\ref{theory}, we give some intuition, why the algorithm works. Section~\ref{conj} conjectures the asymptotic scaling of the SNR based on the numerical results in Section~\ref{num}. Finally, Section~\ref{conc} outlines conclusions for various applications.

We represent scalars, vectors and matrices with non-bold, bold lower case and bold upper case letters, respectively. 
The set of integers is denoted by $\setZ$.
\section{Algorithms}
In the sequel, we will review two algorithms from literature, before introducing the algorithm we propose.
\label{alg}
\subsection{Additive Approximations}
Consider an $N\times K$ matrix $\mM$ that shall be approximated by as few bits as possible.
The standard approach is to approximate $\mM$ by a sum of matrices whose entries are only zeros and ones weighted by scalar factors that are powers of two \cite{zimmermann1999computer,patterson}
\begin{equation}
\label{ssd}
\mM \approx \mA_{\rm sa} = \sum\limits_{q=1}^Q 2^{Q_0-q} \mB_q 
\end{equation}
for $Q_0\in\setZ$ and $\mB_q\in\{0,1\}^{N\times K}$.
Measuring the error by the Frobenius norm $||\mM-\mA_{\rm sa}||_2$, the SNR grows by 6 dB per bit of resolution $Q$.
We will not consider other distortion measures in the sequel.

This approximation requires register shifts for multiplications by powers of two and additions. The additions dominate complexity \cite{zimmermann1999computer,patterson}. Thus, the register shifts are neglected for sake of simplicity.
The standard approach \eqref{ssd} requires $Q/2$ additions per matrix element on average, as no additions are required for the zero entries which occur with probability $\frac12$.

A minor improvement can be achieved, if one allows for more flexibility in the summands.
Let 
\begin{equation}
\label{isd}
\mM \approx \mA_{\rm ia} = \sum\limits_{q=1}^Q \mP_q
\end{equation}
with $\mP_q\in \{0,\pm 2^{\setZ}\}^{N\times K}$.
Since, the complexity of register shifts are of minor importance, we can easily allow for them once per matrix entry. Furthermore, addition and subtraction do not differ in complexity either, so matrix entries may be negative, as well.
For matrices with independent identically distributed (iid) Gaussian entries, this improved approach makes the SNR grow with approximately $\log 27 = 14.3$ dB per bit of resolution $Q$.
However, the probability of zero entries is zero. Thus, we need $Q$ additions per entry of $\mA_{\rm ia}$, twice as many as for $\mA_{\rm sa}$ in the previous approach.
For a single addition per matrix entry, the SNR has, thus, only improved from 12 dB to 14.3 dB. Not really a great deal after all. In both cases, the SNR is independent of the size of the matrix.

Several algorithms in the literature work on additive approximations based on powers of two. Some examples can be followed in \cite{horng1993design,boullis2005some,aksoy2016novel} and the references therein.
\subsection{Multiplicative Approximation}
In this work, we propose the multiplicative approximation
\begin{equation}
\label{mda}
\mM \approx \mA_{\rm m} = \prod\limits_{q=1}^Q \mF_q = \mF_1 \mF_2 \cdots \mF_Q
\end{equation}
with $\mF_1\in \{0,\pm 2^{\setZ}\}^{N\times K}$ and $\mF_q \in \{0,\pm 2^{\setZ}\}^{K\times K}, q>1$. 
Let $K\ge N$ without loss of generality\footnote{If $K<N$ factorize the transpose of the matrix instead.}.
We first obtain $\mF_1$ by quantizing the magnitudes of the entries of $\mM$ to the nearest power of two while keeping the signs untouched.
Thus, $\mF_1=\mP_1$.
To find, the other factors, apply the following recursion:
Approximately factorize the $N\times K$ matrix $\mM$ into $\mM \approx \mL\mR$ such that $\mR$ is square and every column of it is a good solution to the following sparse recovery problem:
\begin{equation}
\label{comsen}
\br_k = \argmin\limits_{\brho\in\{0,\pm 2^{\setZ}\}^K :||\brho||_0=N} || \bm_k - \mL\brho ||_2
\end{equation}
We start the recursion with $\mL = \mF_1$ and $\mR = \mF_2$.
At the $q^{\text{th}}$ recursion, we have
\begin{equation}
\mL = \prod\limits_{i=1}^{q-1}\mF_i \qquad, \qquad \mR = \mF_q.
\end{equation}
Since only $KN$ of the $K^2$ entries of $\mF_2$ to $\mF_Q$ are nonzero, only $Q$ additions per entry of $\mM$ are required.
If this approach is used to directly approximate a square matrix, the performance is poor.
As will become evident in Section~\ref{num}, the algorithm works well if $N$ scales logarithmically in $K$.
In practice, that does not pose a problem unless $N<\log_2 K$, as any large matrix can be split into smaller submatrices which can be approximated independently of each other.
To approximate a $64 \times 64$ matrix, one would decompose it into 6 submatrices of size $6\times 64$ and 4 submatrices of size $7\times 64$.
Then, each of the 10 submatrices is approximated by \eqref{mda}.

There is no reason to set the sparsity in \eqref{comsen} to exactly $N$ other than convenience. In fact, the sparsity in \eqref{comsen} is a free parameter that can be used to trade performance against complexity.

For $Q=1$, the multiplicative and the improved additive approach are identical. 
Thus, we can only hope for improvements by the multiplicative approach for more than one bit per matrix entry unless, we modify the proposed algorithm in a sensible way. Such an improvement is possible by further sparsification and outlined in the sequel:
You may quantize $\mF_1$ harder, setting small elements to zero instead to the closest power of 2. You may also increase the sparsity of $\mF_2$ to $\mF_Q$ by demanding fewer nonzero elements than $N$. This can easily enforced tightening the zero norm constraint in \eqref{comsen}. The precise choice of these parameters are up to optimization and beyond the scope of this paper.
Numerical investigations have shown that in case of further sparsification, the optimum aspect ratio of the submatrices is the less rectangular the further they are sparsified. In fact, it seems to scale in such a way that the relative number of nonzero entries per matrix stays constant.

\section{Numerical Results}
\label{num}
The optimum solution to \eqref{comsen} is NP-hard.
While there are many choices to approximate \eqref{comsen}, we just exemplarily investigate the following decision-directed choice in this section:
Find that component of $\brho$ {such} that, if {we} set {it} to a suitable integer power of two, {the approximation error is reduced most}.
Fix that component to its optimal choice and go on with the second best, third best, etc.\ until you have set $N$ components of $\brho$.

We will restrict the numerical results to matrices $\mM$ that are independent identically distributed (iid) zero Gaussian.
We note, however, that we got very similar results for matrices resulting from optimum linear beam-forming in iid Rayleigh fading.

Table~\ref{rect} shows the resulting SNRs for various matrix sizes and bits of resolution. 
\begin{table}
\begin{center}
\caption{Multiplicative approximation of iid Gaussian random matrices for various levels of resolution $Q$ and sparsity equal to $N$ in \eqref{comsen}.
\label{rect}}
\begin{tabular}{||c||c|c|c|c|c|}
\hline
SNR [dB] & $Q=1$ & $Q=2$ & $Q=3$ & $Q=4$ & $Q=5$\\
\hline\hline
$2\times 4$ &14.2& 20.6 & 24.8 & 27.2 & 28.6\\
$3\times 8$ & 14.2&25.1&32.0&36.3&39.1\\
$4\times 16$ &14.2&   30.0 & 42.1 & 50.7 & 57.1 \\
$5\times 32$ & 14.2&  35.6 &   54.7&  70.7&  82.7 \\
$6\times 64$ &14.2&    41.3&   67.5 &   92.9 &   117\\
$7\times 128$ &14.2 &   47.0&   79.4 &   112&   144\\
$8 \times 256$ &14.2 &  52.6 &   90.8&   129&   167\\
$9 \times 512$ & 14.2 &   58.1 &   102 &   146 &   190 \\
$10 \times 1024$ &14.2&   63.5 &   113 &   162 &   212 \\
$11 \times 2048$ &14.2&   69.1 &   124 &   179 &   234 \\
$12\times 4096$ &14.2&   74.6 &   135 &   195 &   256 \\
$13\times 8192$ & 14.2&80.1& 146 & 212 & 278\\
$14\times 16384$ & 14.2&85.7&157&228&300\\
\hline
\end{tabular}
\end{center}
\end{table}
The results are averaged over at least $10^6$ random variables.
 It is observed that the multiplicative approximation performs poor for small matrices, but performance greatly improves with matrix size.
 For matrices of size $12\times 4096$, any additional level of resolution improves the SNR by slightly more than 60 dB. Comparing at equal number of additions per matrix entry, this is five times the performance of the standard additive approach. 
 Performance can be boosted by further sparsification.

To quantify the benefits of further sparsification, we first introduce the sparsification rate
\begin{equation}
R_r = \frac{||\mF_r||_0}{NK}.
\end{equation}
Without claiming optimality, we set $R_q=R, \forall q$ in the sequel to reduce the number of free parameters.
Table~\ref{fsparse} shows the effects of further sparsification for matrices with $1024$ columns.
\begin{table}
\begin{center}
\caption{Multiplicative approximation of iid Gaussian random matrices of size $10/R \times 1024$ for various sparsification rates $R$ and number of factors $Q$ ranging from 1 to 7.
\label{fsparse}}
{\begin{tabular}{||c||c|c|c|c|c|c|c|c|c||}
\hline
SNR [dB] & $1$ & $2$ & $3$ & $4$ & $5$ & $6$ & $7$\\
\hline\hline
$R=1$ &  14&   64 & 113 & 162 & 212 & 261 & 310\\[1mm]
$R=\frac12$ &   10 &    33 & 55 &     78 & 101 & 123 & 145\\[1mm]
$R=\frac13$ & 6.7&  21&     35 &  49 & 63 &     78 & 92\\[1mm]
$R=\frac14$ & 5.1&  15 &  26&     36& 46 & 55 & 65 \\[1mm]
$R=\frac15$ & 4.2 & 12 & 20 &  28 &   35 & 43 & 50\\[1mm]
$R=\frac16$ & 3.6&  10 &   16 & 23 & 29 &   34 &  40\\[1mm]
$R=\frac17$ & 3.2 & 8.7& 14 & 19 &24 &29 &34\\[1mm]
$R=\frac18$ & 2.9 & 7.6 & 12 & 16 & 21 & 25 & 29\\[1mm]
$R=\frac19$ & 2.6& 6.8 & 11 & 14 & 18 & 21 & 25\\[1mm]
$R=\frac1{10}$ & 2.4 & 6.1& 9.6& 13 & 16 & 19 &22\\[1mm]
$R=\frac1{11}$ & 2.2 & 5.6 & 8.6& 12 & 14 & 17 & 20\\[1mm]
$R=\frac1{12}$ & 2.1& 5.1& 7.9& 11& 13 &15 &18\\[1mm]
\hline
\end{tabular}}
\end{center}
\end{table}
Note from the table that one addition per matrix entry only allows for 14 dB of SNR, while for $R=\frac12$, we already get 33 dB, for $R=\frac13$ even 35 dB. The top SNR is reached for $R=\frac14$ giving 36 dB.
For half an addition per matrix entry, we get 10 dB of SNR for $R=\frac12$, but already 15 dB of SNR for $R=\frac14$.
Sparsification can even help for multiple additions per matrix entry:
Consider two additions: Without sparsification, the SNR is 64 dB, for $R=\frac12$; however, it becomes 78 dB at $Q=4$.
Table~\ref{optsigma} shows some results for optimized sparsification rate.
\begin{table}
\begin{center}
\caption{Multiplicative approximation of iid Gaussian random matrices of various sizes  for optimized sparsification rate vs.\ numbers of additions per matrix entry ranging from $\frac14$ to 3.
\label{optsigma}}
{\begin{tabular}{||l||c||c|c|c|c|c|c|c|c|c|c||}
\hline
SNR [dB] \vphantom{$x^{2^2}$} &$\frac14$& $\frac13$& $\frac12$ & $1$ & $2$ & $3$\\[1mm]
\hline\hline
$K=256$ &6.3&8.3&13&   27 & 60 & 93\\
$K=1024$ &7.9&11&   16 &    36 & 78 &     123\\
$K=4096$ &10&13& 21 &  45&     98 &  152\\
$K=16384$  &12&17 &26 & 55 & 118 & 183\\
\hline
\end{tabular}}
\end{center}
\end{table}

\section{Intuition}
\label{theory}
In this section, we try to provide some preliminary insight why multiplicative approximation gives so good results for large matrices and why the aspect ratio of $\log_2 K\times K$ works so well.
However, we emphasize that we have just started with these investigations and caution the reader that the material in this section is not yet based on solid theory.

\subsection{Combinatorial Approach}

\begin{figure}
\epsfig{file=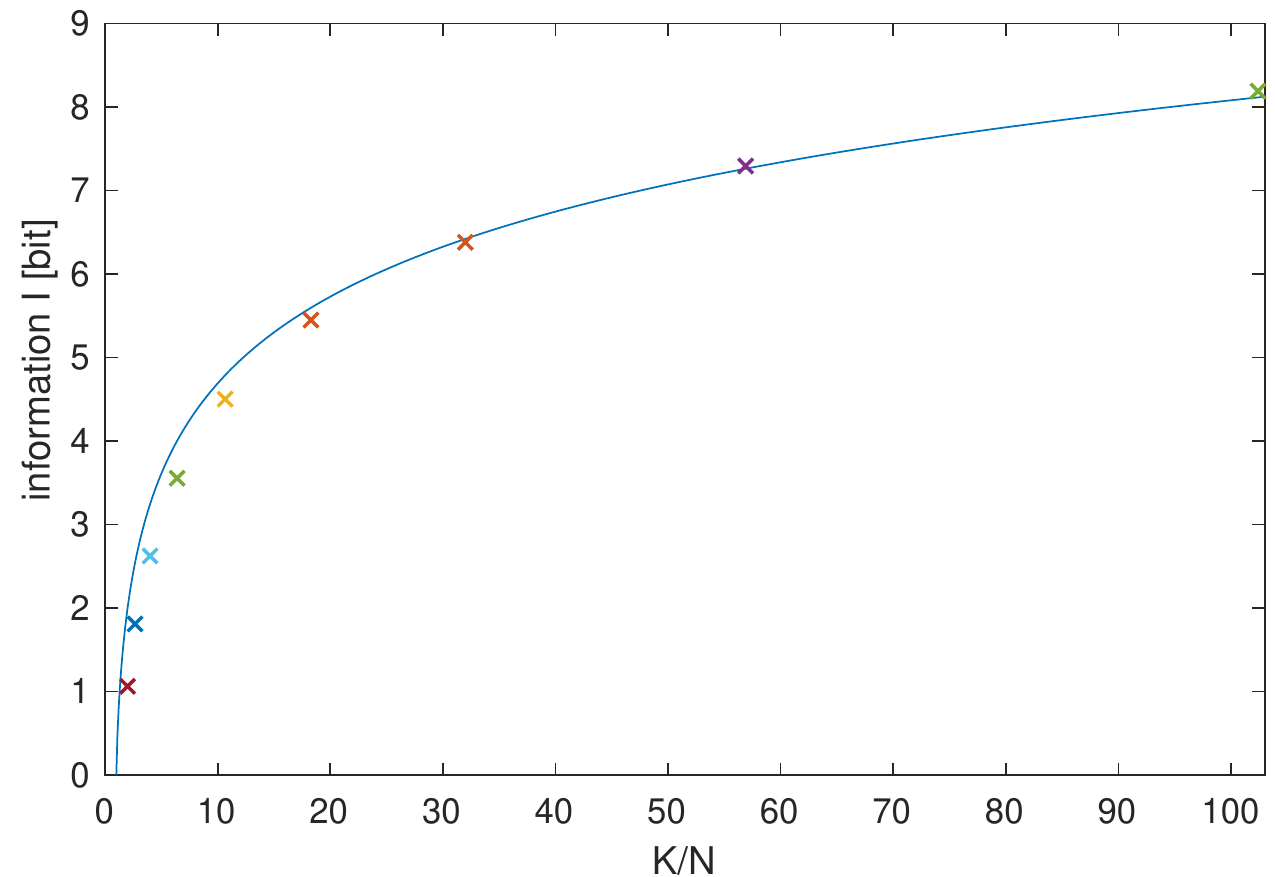,width=\columnwidth}
\caption{Information $I$ from \eqref{eq8} vs.\ aspect ratio. For comparison, the SNR improvements from Table~\ref{rect} are shown by the markers.
\label{fig}
}
\end{figure}

{The} 
information encoded in a matrix approximation is composed of two parts: the precise setting of the powers of two and the positions were the nonzero entries take place. The larger the matrix, the more information is encoded in these positions.
For a 
$K \times K$
matrix with $NK$ nonzero entries, this information is 
\begin{equation}
\log_2\left(
\begin{array}c
K^2\\ NK
\end{array}
\right) \approx K^2 {\rm H}_2 \left(\frac NK\right) \label{eq:Str_approx}
\end{equation}
 where ${\rm H}_2(\cdot)$ denotes the binary entropy function, {i.e.,
\begin{equation}
{\rm H}_2 \left( x \right) = -x \log_2 x - \left(1- x \right) \log_2 \left( 1- x \right)
\end{equation}
for $x\in \left(0, 1 \right)$,} and the right hand side {of \eqref{eq:Str_approx}} follows from Sterling's approximation for the factorial \cite{romik2000stirling}.
Normalizing the information to the number of entries in the matrix to approximate, we get 
\begin{equation}
\label{eq8}
\frac1{NK} \log_2\left(
\begin{array}c
K^2\\ NK
\end{array}
\right) \approx \frac KN  {\rm H}_2 \left(\frac NK\right) =:I.
\end{equation}
The information $I$ is plotted against the aspect ratio {$K/N$} in Fig.~\ref{fig}.
This information is compared against the SNR gains due to increments in $Q$ in Table~\ref{rect}.
For that purpose, the following procedure was used:
For any matrix size, the largest gap between two neighboring columns in Table~\ref{rect} was divided by 6 dB/bit and marked by a cross in the figure. The match is not perfect, but quite close.
This leads us to conjecture that the gains of multiplicative sparse matrix approximations results from the information encoded in the location of nonzero entries.
{As \eqref{eq8} is strictly increasing with the aspect ratio $K/N$, more rectangular matrices are preferred.
\subsection{Geometric Approach}

In the additive approximation \eqref{isd}, the residual error is {bounded from above by} one-third of the approximation calculated in the previous stage. To see this, {note} that the maximum error occurs, if the magnitude of the desired value $v$ is exactly in the middle {of} two adjacent powers of two $a$ and $2a$.
{In this case}, we have 
\[
|v| -a = 2a - |v| \qquad \Rightarrow \qquad |v|-a = \frac{|v|}3.
\]
Assuming the {residual} error to be uniformly\footnote{Calculating the error exactly with respect to its true slightly non-uniform distribution, that arises from the Gaussianity of the source, leads to deviations around 0.1 dB.}  distributed, it is easily {concluded that the error has} power $v^2/27$. Thus, every additional summand {improves the SNR by a factor of 27.}

In the multiplicative case, we approximate the $N$-dimensional target vector $\bv$ by scaling the best suited one out of the $K$ base vectors provided by the columns of $\mL$ in \eqref{comsen} with a power of two.
We take the residual error vector as the new target vector and repeat the procedure for a total of $NR$ times per matrix factor.
Two types of errors are possible {in this case, namely missing} the target by direction and by distance.
By the orthogonality principle, these two errors are orthogonal to each other, cf.\ Fig.~\ref{pic}.
\begin{figure}
\centerline{
\epsfig{file=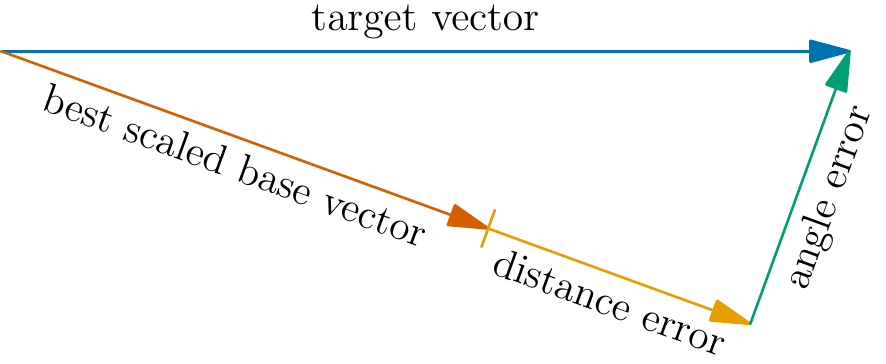,width=.8\columnwidth}}
\caption{\label{pic} Decomposition of the approximation error.}
\end{figure}
Denoting the angle between the target vector $\bv$ and the best base vector by $\alpha$, the two errors become $\left\Vert \bv \right\Vert \left\vert\sin\alpha \right\vert$ and $\left\Vert \bv \right\Vert \left\vert\cos\alpha \right\vert a$, {respectively, where} $a$ {is assumed} uniformly distributed on} $[0 ;\frac13)$.

In order to approximate the angle $\alpha$, we assume a logarithmic relation between the number of rows and columns, i.e.\
\begin{equation}
\label{logscale}
N=\frac1R  \log_ 2 K.
\end{equation} 
We normalize all base vectors and the target vector without loss of generality to unit length.  Now, they all lie on an $N$-dimensional hypersphere, whereon we assume them to be uniformly distributed. Without loss of generality, we set the target vector to be the first unit vector of the Carthesian coordinate system.

The uniform distribution on the hypersphere is canonically created by normalizing an iid Gaussian random vector to unit norm.
Let $\varrho_k$ be the magnitude of the crosscorrelation coefficient between the target vector and the $k$-th base vector. Then, we have
\begin{equation}
\cos\alpha = \max\limits_k \varrho_k.
\end{equation}
We can now construct the squared correlation coefficient $\varrho_k^2$ out of $N$ iid Gaussian random variables $g_n$ as
\begin{equation}
\label{rayleigh}
{\varrho}_k^2 = \frac{{g_1^2} }{{\sum\limits_{n=1}^N g_n^2}}.
\end{equation}
In the numerator, only the first Gaussian random variable shows up due to the inner product with the first unit vector of the coordinate system.
The ratio in \eqref{rayleigh} is known to be distributed according to the beta distribution with shape parameters $\frac12$ and $\frac {N-1}2$. The corresponding density is given by \cite{casella2002statistical}
\begin{equation}
p_{\varrho^2}(r) = \frac1Z  r^{-\frac12} (1-r)^{\frac {N-3}2}.
\end{equation}
{Here, $Z=B(\frac12,\frac {N-1}2)$} is a normalizing constant, also known as \textit{partition function}, with $B(\cdot,\cdot)$ denoting the Beta function.

Substituting ${\varrho}=\sqrt r$ leads to the density
\begin{align}
p_{\varrho}(\varrho) &= \frac2Z  \left(1-\varrho^2\right)^{\frac {N-3}2}
\end{align}
and the cumulative distribution function
\begin{equation}
P_{\varrho}(\varrho) = \frac 2Z \int\limits_0^\varrho (1-\xi^2)^{\frac {N-3}2} {\text d}\xi.
\end{equation}
The distribution of the maximum
\begin{align}
P_{\max}(\varrho) = \left[P_{\varrho}(\varrho)\right]^K
\end{align}
is shown in the Appendix to converge to
\begin{equation}
\lim\limits_{K\to\infty} \lim\limits_{N\to\frac1R\log_2 K} P_{\max}(\varrho) = \left\{
\begin{array}{ll}
0 & \varrho < \sqrt{1-4^{-R}}\\[1mm]
1 & \varrho > \sqrt{1-4^{-R}}
\end{array}
\right..
\end{equation}
under the scaling law \eqref{logscale}.
Thus, we find
\begin{equation}
\cos^2\alpha = 1-4^{-R}\qquad ,\qquad \sin^2\alpha = 4^{-R}.
\label{sine}
\end{equation}
{Successive reductions of approximation errors multiplicatively compound on top of previous reductions}.
Thus, we need to average with respect to the distance error in the logarithmic domain.
For $R=1$, the expected SNR becomes
\begin{equation}
\gamma= \exp \left[ -3 \int\limits_0^{\frac 13} \ln \left( \frac14 + \frac 34 a^2 \right) {\text d} a\right] = 3 \exp\left(2-
\frac{\sqrt 3 \pi}3 \right)
\end{equation}
which is well approximated by 3.614.
It hardly differs from $\frac {18}5=3.6$ which arises from linear instead of logarithmic averaging over the distance error.

Finally, the SNR per matrix factor becomes
\begin{align}
\gamma ^{NR}
\end{align}
since the projection procedure is repeated for any of the $NR$ nonzero entries of the rows of the right matrix factors.
Note that the SNR grows unboundedly with matrix size $N$.
Furthermore, the error in direction dominates over the error in distance. Thus, there is no point in going for multiple bits of resolution per matrix factor.

\subsection{Optimum Aspect Ratio}
The aspect ratio cannot become arbitrarily large. If the number of rows $N$ becomes too small, the probability that two columns become identical rises. In the compressive sensing interpretation of \eqref{comsen}, identical columns mean to take the same measurement twice. This is clearly a waste of resources.

Let $p$ be the probability that two matrix entries quantized to powers of two are identical. Then, the probability that two columns of $\mF_1$ are identical is $p^N$. Since there are $K(K-1)/2$ different pairs of columns in total, {we can use the union bound to show that the probability of having at least two identical columns is smaller than}
\begin{equation}
p^N \frac{K(K-1)}2.
\end{equation}
If we set a fixed threshold for that event to happen and solve for $N$, we see that it scales logarithmically in $K$. 

\subsection{Connection to Rate-Distortion Theory}
The sparsification rate $R$ has an interesting interpretation {by looking at the problem from a rate-distortion theory point of view}. Consider a Gaussian source being lossy compressed to rate $R$. {Following the quadratic Gaussian source coding \cite{el2011network}}, {the expression derived for} the squared angle error in \eqref{sine}, {i.e., $4^{-R}$,} gives the mean-squared distortion achieved {for this compression rate. In other words, by}  interpreting the columns of matrix $\mL$ in \eqref{comsen} {as entries of} a Gaussian {codebook} {and sparsification rate $R$ as its compression rate, the squared angle error {in Fig.~\ref{pic}} corresponds to the distortion specified by the rate-distortion curve.}

\section{Outlook}
\label{conj}
For optimum aspect ratio, the numerical results suggest that the SNR scales as
\begin{equation}
\left(
\frac{RK^2}{\log_2 K}
\right)^{(Q-1)R}
\end{equation}
Note that this does not only grow to infinity for large $Q$, but also for large $K$ whenever $Q>1$.
Note that the number of additions per matrix entry is $QR$. Thus, for $K$ growing large and $R$ vanishing with $1/R ={\text o}(\log K)$, the computational effort per matrix entry goes to zero while, at the same time, the SNR goes to infinity.
\section{Conclusions}
\label{conc}
Matrix multiplications can be implemented much more efficiently by multiplicative decompositions, if the matrix is sufficiently large.
For applications in massive MIMO and millimeter wave communications, beamforming {can eventually be implemented} with even less than half an addition per matrix entry. However, these gains come at the expense of larger storage requirements for the sparse matrices.
Neural networks, however, may be hard wired after having been trained. In this case, no additional storage is required and the gains can be utilized in full. In field programmable gate array (FPGA) implementations even an update of the factorized matrix is possible. Assuming that our findings generalize to nonlinear multidimensional functions, neural networks need neither have precise weights nor be densely connected, if they are sufficiently deep. Any quantization errors at intermediate layers and any level of sparsity can be compensated for by more layers and larger dimensions.

\section*{Appendix}
In order to show the convergence of the cumulative distribution function of the maximum correlation coefficient to the unit step function, recall the following limit holding for any positive $x$ and $r$
\begin{equation}
\lim\limits_{K\to\infty}\left(
1- \frac x{K^r}
\right)^K = \left\{
\begin{array}{ll}
0 & r<1\\
\exp(x) & r=1\\
1 & r >1
\end{array}
\right..
\end{equation}
The limiting behavior of $P_{\max}(\varrho)$ is, thus, decided by the scaling of $1-P_{\varrho}(\varrho)$ with respect to $K$. The critical scaling is $\frac 1K$.
Such a scaling implies a slope of $-1$ in doubly logarithmic scale.
Thus,
\begin{equation}
\lim\limits_{N\to\infty} \frac{\partial}{\partial (NR)}
\log_2 \left[1 - P_{\varrho}(\varrho)\right]  = -1.
\end{equation}
Explicit calculation of the derivative yields
\begin{equation}
\lim\limits_{N\to\infty}
\frac{
\int\limits^1_\varrho (1-\xi^2)^{\frac {N-3}2} 
 \log_2\left(1-\xi^2\right)
{\text d}\xi.
}{2R
\int\limits^1_\varrho (1-\xi^2)^{\frac {N-3}2} {\text d}\xi.
}=-1
\end{equation}
and saddle point integration gives \cite[Chapter 4]{merhav2010statistical}
\begin{equation}
\frac 1{2R} \log_2\left(1-\varrho^2\right) =-1.
\end{equation}
This immediately leads to $\varrho^2 =1-4^{-R}$.

\bibliography{ref}
\bibliographystyle{IEEEtran}
\end{document}

%% file: commands.tex
\newcommand{\setZ}{\mathbb{Z}}

\newcommand{\br}{{\mathbf{r}}}

\newcommand{\bv}{{\boldsymbol{v}}}
\newcommand{\bm}{{\boldsymbol{m}}}
\newcommand{\brho}{{\boldsymbol{\rho}}}

\DeclareMathOperator*{\argmin}{\mathrm{argmin}\hspace*{1mm}}

\newcommand{\mA}{\mathbf{A}}
\newcommand{\mB}{\mathbf{B}}

\newcommand{\mR}{\mathbf{R}}

\newcommand{\mP}{\mathbf{P}}
\newcommand{\mL}{\mathbf{L}}
\newcommand{\mF}{\mathbf{F}}

\newcommand{\mM}{\mathbf{M}}

\newtheoremstyle{mystyle}
  {}
  {}
  {}
  {}
  {\bfseries}
  {:}
  { }
  {}
\theoremstyle{mystyle}

\algnewcommand\algorithmicLet{\textbf{Let}}
\algnewcommand\Let{\item[\algorithmicLet]}
\algnewcommand\algorithmicSet{\textbf{Set}}
\algnewcommand\Set{\item[\algorithmicSet]}

\algnewcommand\algorithmicInitiate{\textbf{Initiate}}
\algnewcommand\Initiate{\item[\algorithmicInitiate]}
\algnewcommand\algorithmicStart{\textbf{Begin}}
\algnewcommand\Begin{\item[\algorithmicStart]}
\algnewcommand\algorithmicEnd{\textbf{End}}
\algnewcommand\End{\item[\algorithmicEnd]}

\algnewcommand\algorithmicOutP{\textbf{Output:}}
\algnewcommand\Out{\item[\algorithmicOutP]}

\algnewcommand\algorithmicInP{\textbf{Input:}}
\algnewcommand\In{\item[\algorithmicInP]}

\newcounter{bar}

